\documentclass[aip, jap, reprint, groupedaddress, 10pt]{revtex4-1}
\usepackage[utf8]{inputenc}
\usepackage{helvet, graphicx}
\usepackage{hyperref} 
\usepackage[usenames]{color}
\usepackage{amsmath,amssymb,amsfonts, braket}
\usepackage{bm}
\usepackage[margin=0.75in]{geometry}
\usepackage[T1]{fontenc}

\usepackage{setspace}

\renewcommand{\figurename}{FIG.}

\newcommand{\er}{Er$^{3+}$ }
\newcommand{\upg}{\uparrow_\text{g}}
\newcommand{\downg}{\downarrow_\text{g}}

\renewcommand{\thesection}{\Roman{section}} 
\renewcommand\thesubsection{\Alph{subsection}}

\begin{document}
\title{Hybrid microwave-optical scanning probe for addressing solid-state spins in nanophotonic cavities}

\author{Songtao Chen}
\author{Salim Ourari}
\author{Mouktik Raha}
\author{Christopher M.~Phenicie}
\author{Mehmet T.~Uysal}

\author{Jeff D.~Thompson}
\thanks{Author to whom correspondence should be addressed: jdthompson@princeton.edu}
\affiliation{Department of Electrical Engineering, Princeton University, Princeton, NJ 08544, USA\looseness=-1}

\date{\today}

\begin{abstract}
Spin-photon interfaces based on solid-state atomic defects have enabled a variety of key applications in quantum information processing. To maximize the light-matter coupling strength, defects are often placed inside nanoscale devices. Efficiently coupling light and microwave radiation into these structures is an experimental challenge, especially in cryogenic or high vacuum environments with limited sample access. In this work, we demonstrate a fiber-based scanning probe that simultaneously couples light into a planar photonic circuit and delivers high power microwaves for driving electron spin transitions. The optical portion achieves 46\% one-way coupling efficiency, while the microwave portion supplies an AC magnetic field with strength up to 9 Gauss. The entire probe can be scanned across a large number of devices inside a $^3$He cryostat without free-space optical access. We demonstrate this technique with silicon nanophotonic circuits coupled to single \er ions.
\end{abstract}

\maketitle

\section{INTRODUCTION}
\label{sec:intro} 
Optically interfaced solid-state spins are promising platforms for quantum technologies \cite{awschalom2018}. In particular, spin-photon entanglement \cite{togan2010, gao2012} has enabled deterministic entanglement of remote spins \cite{bernien2013} and the realization of spin-photon gates \cite{bhaskar2020experimental, sun2016quantum}. To maximize light-matter interactions, nanophotonic cavity structures are widely exploited in different solid-state spin systems, such as GaAs quantum dots \cite{englund2005controlling, badolato2005deterministic}, color centers in diamond \cite{sipahigil2016integrated} and SiC \cite{crook2020}, and rare earth ions \cite{miyazono2016coupling, dibos2018, zhong2018optically}.

Several techniques have been demonstrated for coupling light from an optical fiber into a nanophotonic waveguide or cavity. Tapered optical fibers have realized coupling efficiencies as high as 97\% through adiabatic mode transfer \cite{tiecke2015efficient} between the fiber and waveguide. Another approach is end-fire coupling, using cleaved or lensed optical fibers to couple light directly into the end facet of a waveguide \cite{cohen2013optical, meenehan2014silicon, dibos2018}. Lastly, grating couplers (GC) can extract light from planar waveguides and radiate it out of the plane and into a fiber above the sample surface \cite{halir2015,marchetti2019coupling}.

In addition to optical access, an intense microwave (MW) magnetic field is necessary to drive electron spin transitions. A common approach to generate the field is to use the lithographically patterned striplines on the sample surface \cite{toyli2010chip}, connected to external driving circuits using wirebonds or solder.

In many cases, it is desirable to be able to produce an array of devices and realize optical and MW coupling to any device in the array. One example case is to probe a set of devices with a swept parameter. In the techniques discussed above for coupling light from a fiber to a waveguide, the fiber can be moved using a nanopositioner to couple to devices across a large-scale array. However, the connections to MW circuits are permanent and fixed, and coupling to a large number of MW devices may require many wirebonds and coaxial feedthroughs which is a significant practical limitation.

Here, we present a new approach to realize simultaneous optical and MW coupling to devices across a large-scale array using a scanning probe (Fig.~\ref{figure1}a). The probe has an optical portion that couples light from a fiber into a nanophotonic waveguide using a GC \cite{snyder2013,mckenna2019}, and a MW coplanar waveguide (CPW) that delivers intense MW magnetic fields at the sample surface. This approach enables uniform optical coupling efficiency and MW drive strength for any device in the array, with only a single optical and MW connection to the probe head. The probe is capable of working inside a $^3$He cryostat at 540~mK, and can be aligned and scanned between devices without requiring free-space optical access. In this paper, we discuss the design and fabrication of this device, and demonstrate an application to observing and manipulating single \er ions in a Y$_2$SiO$_5$ (YSO) crystal coupled to a silicon nanophotonic cavity \cite{dibos2018,raha2020,chen2020}.

\section{Device design and fabrication}
\label{sec:probe} 
\begin{figure}[!t]
	\centering
    \includegraphics[width = 84.790 mm]{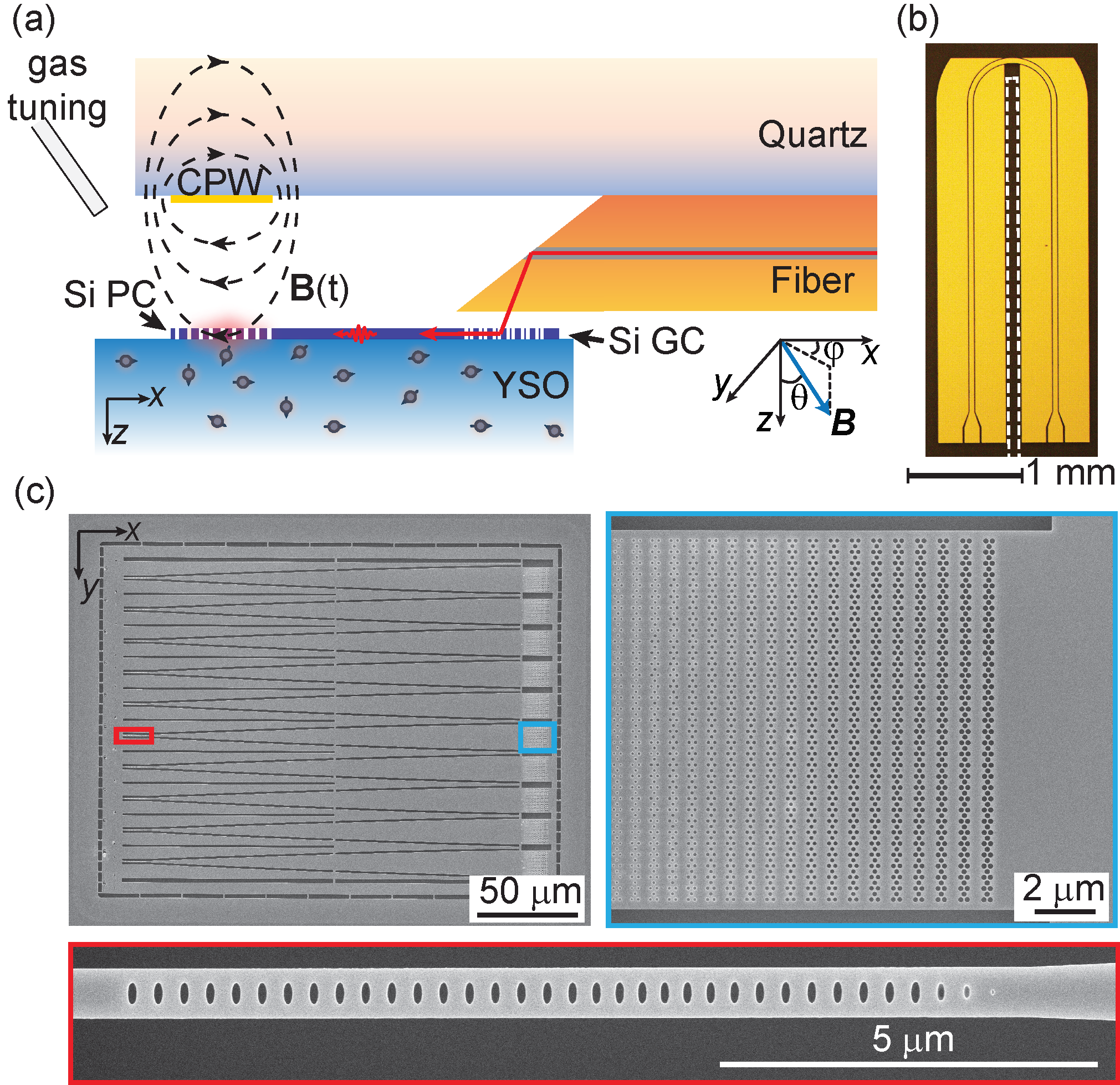}

    \caption{ \textbf{MW-optical hybrid scanning probe}.
    \textbf{(a)} A schematic illustration of the device assembly.
    \textbf{(b)} Optical image of the CPW structure on the quartz surface. The angle-polished fiber is glued between two inner ground pads (inside the white dashed box).
    \textbf{(c)} Scanning electron microscope images of the silicon nanophotonic circuits that couple to the probe. A device block consists of ten individual devices, each of which has a PC cavity (red rectangle) and a GC (blue rectangle) connected by an adiabatic waveguide taper. 
    } 
	\label{figure1}
	\end{figure}

A schematic of the hybrid probe is depicted in Fig.~\ref{figure1}a. The probe consists of a lithographically patterned CPW structure (Fig.~\ref{figure1}b) and an angle-polished fiber glued on a quartz substrate. Light is totally internally reflected at the polished fiber surface \cite{snyder2013,mckenna2019} and directed downwards to the sample. The entire probe is mounted on a three-axis piezo stage (Attocube Inc.). The nanophotonic structures to which we want to couple are fabricated on a SOI wafer and transferred onto the YSO crystal doped with \er ions via a stamping technique\cite{dibos2018}. Light emitted by the ions couples to a photonic crystal (PC) cavity (Fig.~\ref{figure1}c, red rectangle) at one end, and an adiabatic waveguide carries the light to a GC (Fig.~\ref{figure1}c, blue rectangle) that scatters it up and into the fiber. The minimum distance between the cavity and GC is imposed by the adiabaticity of the waveguide taper (185 $\mu$m in length). The spacing between the fiber and CPW is chosen to match the waveguide taper length, placing the CPW directly above the atomic defects that couple to the cavity, to maximize the MW field strength. The front edge of the CPW substrate is diced $\sim$ 10 $\mu$m from the CPW center pin so as not to block the pathway for nitrogen gas condensation (Fig.~\ref{figure1}a), which is used to fine tune the cavity resonance at cryogenic temperatures \cite{mosor2005scanning}.

\subsection{Grating coupler}
\label{sec:GC}
	\begin{figure*}[t]
	\centering
    \includegraphics[width = 160.147 mm]{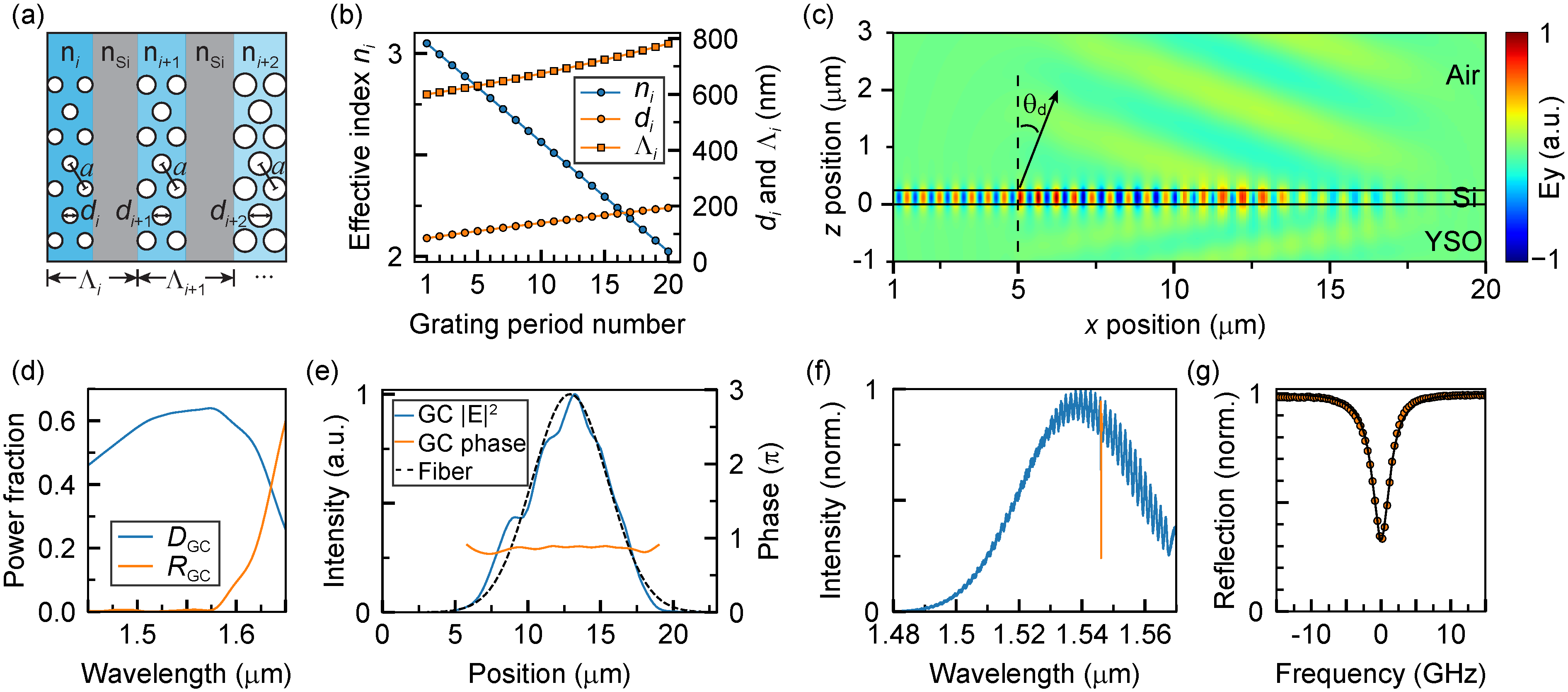}
        
    \caption{ \textbf{2D SWG grating coupler}.
    \textbf{(a)} Schematic view and structural parameters of the GC. Here, $n_i$, $\Lambda_i$ and $d_i$ are the effective index, pitch and photonic crystal hole diameter in each grating period. The photonic crystal lattice constant is $a$, and $n_{\textrm{Si}}$ denotes the refractive index of silicon.
    \textbf{(b)} Parameters of the optimized, apodized grating structure.
    \textbf{(c)} Simulated electric field ($E_\text{y}$) radiated by the GC. $\theta_\text{d}$ denotes the diffraction angle.
    \textbf{(d)} Simulated $D_\text{GC}$ and $R_\text{GC}$ as a function of wavelength.
    \textbf{(e)} Simulated intensity and phase profiles of the GC diffracted field, together with the mode profile of an SMF-28 fiber.
    \textbf{(f)} Measured coupling to a PC cavity device, with the cavity resonance shown in orange.
    \textbf{(g)} Narrow scan of the cavity showing quality factor Q = 6$\times$10$^4$.
    }
	\label{figure2}
	\end{figure*}

First, we discuss the GC and angle-polished fiber used to couple light into the PC cavity. Grating couplers have been widely used to interface nanophotonic waveguides to far-field radiation modes and single-mode optical fibers \cite{halir2015,marchetti2019coupling}. Uniform gratings do not provide good mode matching to Gaussian beams or optical fiber modes \cite{halir2015}; realizing better mode matching requires apodizing the grating strength in some manner. One approach is partially-etched grating structures, where the grating strength is controlled by the partial etching depth \cite{li2013}. However, this method requires multiple etching steps in the fabrication. Instead, we pursue single-step etched GCs based on 2D photonic crystal subwavelength gratings (SWG) \cite{liu2010, ding2013}, where the grating strength can be modulated by changing the effective index of the photonic crystal.

A schematic of the GC is shown in Fig.~\ref{figure2}a. To shape the spatial profile of the GC-diffracted field, the effective index $n_i$ of each grating period is modulated by controlling the hole diameter ($d_i$) of a triangular 2D photonic crystal (with constant pitch $a=215$ nm) \cite{liu2010, ding2013}. The grating periodicity ($\Lambda_i$) is tuned simultaneously to ensure a constant diffraction angle by matching the phase of the guided light with the diffracted light. The subwavelength nature of the SWG structure ($a < \lambda$) ensures that it acts as an effective medium without introducing extra parasitic in-plane diffraction \cite{halir2010, benedikovic2014}. Both the GC and the PC cavity use the transverse-electric waveguide mode (with predominant polarization along $\hat{y}$).

The coupling efficiency of the GC can be written as $\eta = D_\text{GC}O_\text{GC}$, where $D_\text{GC}$ is the directionality (the fraction of the incident power in the waveguide that is directed upwards towards the fiber), and $O_\text{GC}$ is the overlap of the diffracted mode with the single mode fiber. The fiber overlap can be further separated into a component along the grating direction $O_{\text{GC},x}$, determined by the grating properties, and a component in the perpendicular direction, $O_{\text{GC},y}$, which is matched by choosing the waveguide width $w_y=12.6\,\mu$m (for SMF-28 fiber). $D_\text{GC}$ is influenced by the amount of light directed down into the sample as well as light reflected back into the waveguide, $R_\text{GC}$. In addition to reducing $\eta$, $R_\text{GC}$ also results in interference fringes because of standing waves between the cavity and the GC.

The design of the grating proceeds in several steps, following Ref. \citenum{halir2009waveguide}. First, a 2D FDTD simulation ($xz$-plane) of a non-apodized grating consisting of alternating slabs of fixed indices $n_\text{d}$ and $n_\text{Si}$ is used to optimize $D_\text{GC}$ by varying $n_\text{d}$, the periodicity, $\Lambda$, and the duty cycle, $D$. We find optimum values $\Lambda=600$ nm and $D=0.5$, which are used as a starting point for the first grating period in the apodized structure. Then, a set of ratios between the hole diameter and the photonic crystal lattice constant $d_i/a$ are found analytically to realize various $n_\text{eff}$ \cite{ding2013}. Finally, the apodized grating is simulated using a 3D FDTD simulation\cite{halir2009waveguide} (with periodic boundary conditions along $\hat{y}$) to determine all of the terms contributing to $\eta$ (except $O_{\text{GC},y}$). This simulation is repeated for various linear gradients in $n_i$, and we find that $\eta$ is maximized for a first grating period with index $n_1 = 3.05$ and a step between periods of $\Delta n = -0.054$ (Fig.~\ref{figure2}b). The resulting diffraction angle is $\theta_d = 11.3^\circ$, and the electric field distribution is shown in Fig.~\ref{figure2}c.


The simulated GC achieves $D_\text{GC}$ = 62.5\% and $R_\text{GC}$ = 0.5\% at $\lambda = 1536$ nm (Fig.~\ref{figure2}d). The diffracted field has a nearly constant phase profile and features a 97.7\% mode overlap with the fiber in the $\hat{x}$-direction (Fig.~\ref{figure2}e). After considering the transmission of the adiabatic waveguide taper (96.2\%), $O_{\text{GC},y}$ (98.3\%), and the glass-air interface reflection loss (3.6\%), we estimate a total simulated efficiency of $\eta_\text{sim}$ = 55.7\%.

To experimentally characterize the GC performance, we couple an angle-polished fiber to the device and measure the reflection signal $R$ from the PC cavity (away from the cavity resonance), and estimate the one-way coupling efficiency \cite{tiecke2015efficient} according to $\eta=\sqrt{R}$. We use an automatic fiber polisher (ULTRA TEC.) to polish the fiber end with an angle $\theta_p = 41.2^\circ$, such that the reflected beam is incident onto the GC at the same angle $[\sin^{-1}(n_\text{fiber}\sin(\pi/2 - 2\theta_p))$, $n_\text{fiber} = 1.47]$ as the simulated GC diffraction angle $\theta_d$. Measurements across multiple stamped devices give an average $\eta$ = 46.2 $\pm$ 2.6\%. The coupler 1 dB bandwidth is 33 nm (Fig.~\ref{figure2}f), limited by the large effective waveguide mode index of the gratings \cite{halir2015, marchetti2019coupling}. The pronounced interference fringes ($\sim$ $\pm$5\%) observed at longer wavelengths (Fig.~\ref{figure2}f) result from  Fabry-Perot (FP) modes generated between the GC and PC cavity. The estimated FP cavity length roughly matches with the length of the adiabatic waveguide taper, and the wavelength dependence of the oscillation amplitude matches the simulated $R_{\text{GC}}$ (Fig.~\ref{figure2}d). Despite these oscillations, the coupling efficiency is flat over the linewidth of a moderately high-Q cavity (Fig.~\ref{figure2}g). 


\subsection{Microwave coplanar waveguide}
\label{sec:CPW}
Next, we discuss the MW component of the scanning probe. Microwave striplines or coplanar waveguides are commonly used to deliver intense microwaves for spin resonance experiments, as they can concentrate the energy density into a small area to generate high field intensities. In many experiments, the CPW is patterned directly on the sample surface, which has the benefit of allowing extremely small separation between the spins and the CPW. However, it has the disadvantage of requiring extra fabrication steps and wirebonding, and it is difficult to connect a large number of devices. Here, we pursue an alternate approach: we fabricate a CPW as part of the scanning probe. This enables addressing spins coupled to different optical cavities while we scan the probe position, and does not require fabrication on the sample itself.

\begin{figure*}[t]
	\centering
    \includegraphics[width = 149.7838 mm]{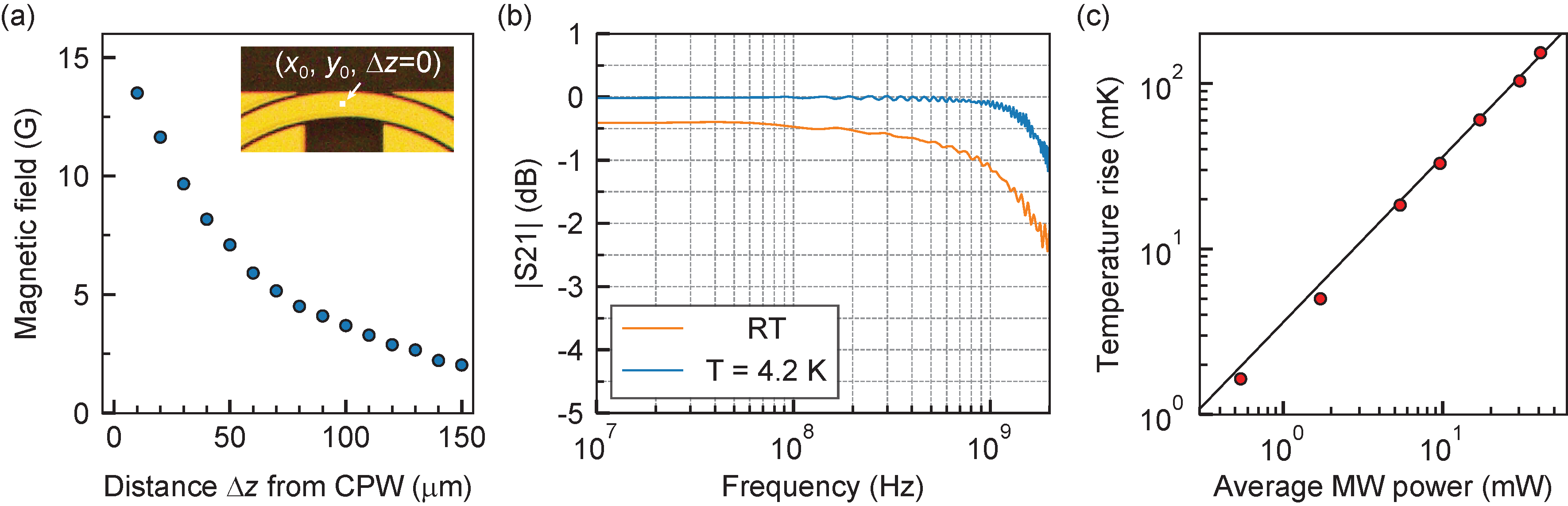}
    
    \caption{ \textbf{Microwave structure design and characterization}.
    \textbf{(a)} Simulated AC magnetic field strength (1 Watt input power, $f_\text{MW}$ = 1.76 GHz) at varying distance $\Delta z$ from the CPW surface.
    \textbf{(b)} Transmission (|S21|) of CPW and PCB at room temperature (RT) and T = 4.2 K.
    \textbf{(c)} Heating of the lowest temperature stage in the cryostat with CW MW input at varying power levels (red dots), and a linear fit with slope 3.6 mK/mW.
    } 
	\label{figure3}
	\end{figure*}

We turn to MW field simulation to determine the optimum CPW structural parameters. The designed gold CPW structure (Fig.~\ref{figure1}b) has center pin and gap widths of 50 $\mu$m and 5 $\mu$m respectively, which produces large field strengths at the intended separation from the CPW surface. With 1 Watt of input power, we calculate a field of ($B_\text{AC},~\theta,~\varphi$) = (2.8 G, 69.2$^\circ$, 1.2$^\circ$) at the closet point on the sample surface, 125 $\mu$m away from the CPW center pin (Fig.~\ref{figure3}a). The gap in the ground plane was introduced to make the fiber tip visible from the backside for alignment, although this turned out not to be necessary and a full ground plane could be restored.

The fabricated CPW consists of a 2-inch quartz wafer (MTI Corp.) with a lithographically patterned gold layer (1 $\mu$m in thickness) deposited by electron beam evaporation. Individual quartz carrier chips (4.1 $\times$ 1.7 mm$^2$) with CPWs are then diced from the wafer, with one of the cuts made as close as possible to the CPW to avoid obstructing the gas condensation pathway for cavity tuning. The CPW is wirebonded to a carrier PCB, which is mounted to the piezo stage. SMA connectors on the PCB connect to MW feedthroughs (both input and output), which are heat sunk at each stage of the cryostat using 0 dB attenuators.

Next, we characterize the insertion loss of the CPW and its heat load on the cryostat. The insertion loss of CPW and PCB, measured by a network analyzer with the cable losses subtracted, is less than 1 dB up to $\approx2$ GHz (Fig.~\ref{figure3}b). After installation in the cryostat (Bluefors LD-250-HE), the total insertion loss at $1.76$ GHz is 4.7 dB, which is dominated by cupronickel coaxial cables between the 4K and 500 mK stages on both the input and output lines. When significant MW power is applied, the lowest temperature (500 mK) stage of the cryostat is heated by MW absorption. We measure an increase in temperature on the 500 mK stage of 3.6~mK/mW of absorbed MW power (Fig.~\ref{figure3}c). The measured cooling power of this stage (0.07~mW/mK) would result in nearly four times greater temperature increase, indicating that a significant fraction of the absorbed power is dissipated at the 4K stage with much higher cooling power. In pulsed electron spin resonance experiments, we have used short pulses with up to 20~W peak power and observed no difference in the heating rate from a CW MW tone with the same average power. While the cable losses could be reduced considerably in principle by using superconducting cables between the 4K and 500 mK stages, it is unclear whether the peak AC currents (0.9 A for 20 W) would exceed the critical current of these cables.





\subsection{Fiber alignment and device assembly}
\label{sec:fiber}

\begin{figure}[!b]
	\centering
    \includegraphics[width = 84.015 mm]{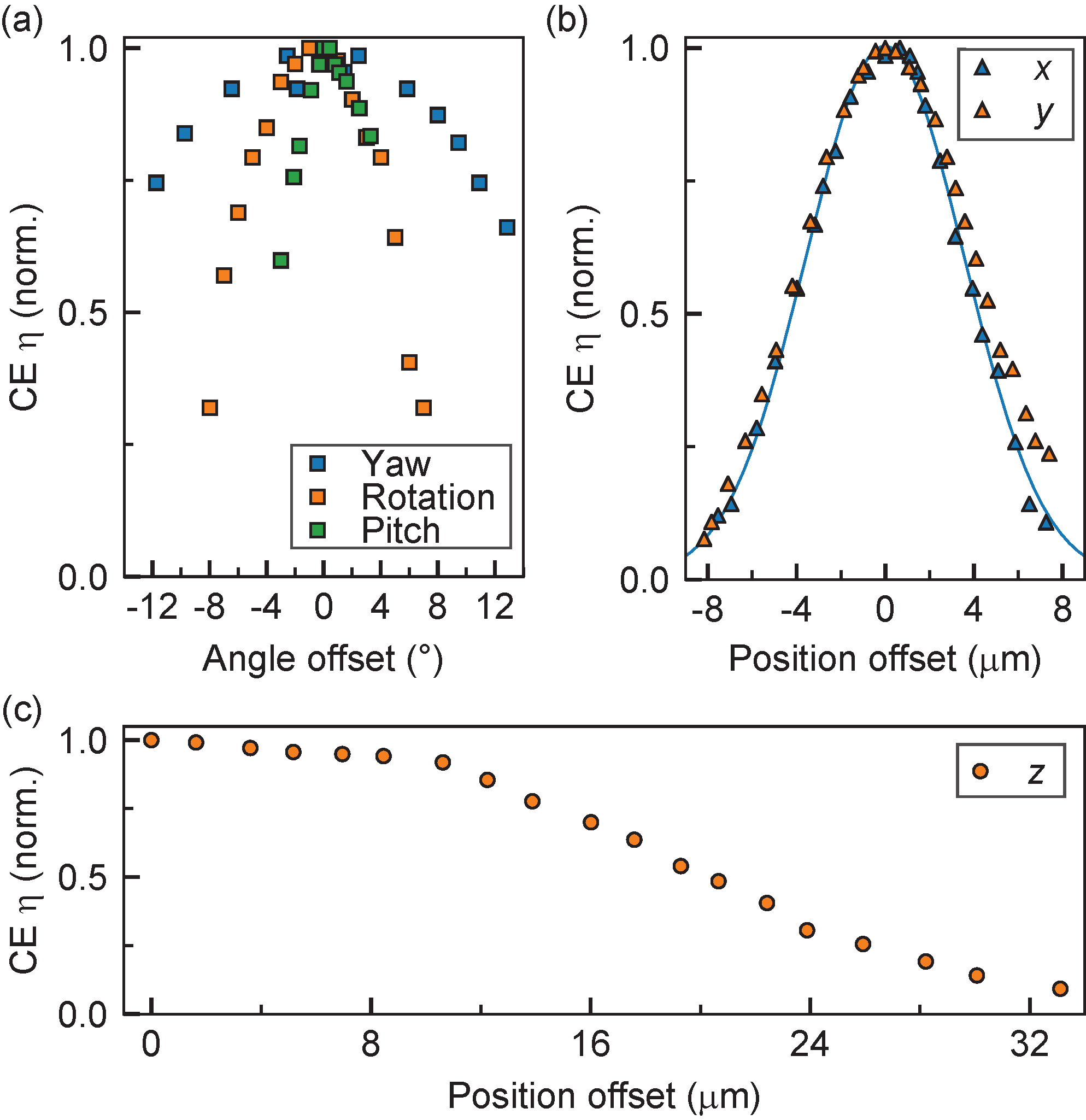}

    \caption{\textbf{Fiber alignment tolerances}.
    \textbf{(a)} Angle dependence of $\eta$. At each angular offset, the translation is re-optimized to remove the coupling between these coordinates in the translation stage.
    \textbf{(b)} In-plane position dependence of $\eta$, together with a Gaussian profile with $1/e^2$ diameter 14.3 $\mu$m.
    \textbf{(c)} Out-of-plane position dependence of $\eta$.
    }
	\label{figure4}
	\end{figure}

Lastly, we consider assembly of the angle-polished fiber and the CPW. To determine the accuracy required for this assembly, we investigate the sensitivity of the fiber-device coupling efficiency to misalignment. The coupling efficiency (CE) $\eta$ is affected by all six degrees of freedom: the yaw, rotation and pitch angles (Fig.~\ref{figure4}a) and spatial offsets in $\hat{x}$, $\hat{y}$ and $\hat{z}$ (Fig.~\ref{figure4}b,c). We find that the efficiency is most sensitive to pitch and rotation errors, which strongly affect the mode overlap ($O_\text{GC}$). Achieving coupling within 95\% of the maximum requires an alignment accuracy of $\pm$ 1$^{\circ}$. The $\eta$-dependence on the transverse positions ($\hat{x}$, $\hat{y}$) reveals a Gaussian profile requiring a tolerance of $\pm$ 1 $\mu$m to maintain 95\% of the maximum coupling. In the $\hat{z}$-direction, $\eta$ decreases monotonically with the distance from the GC. 

	\begin{figure*}[t]
	\centering
    \includegraphics[width = 151.630 mm]{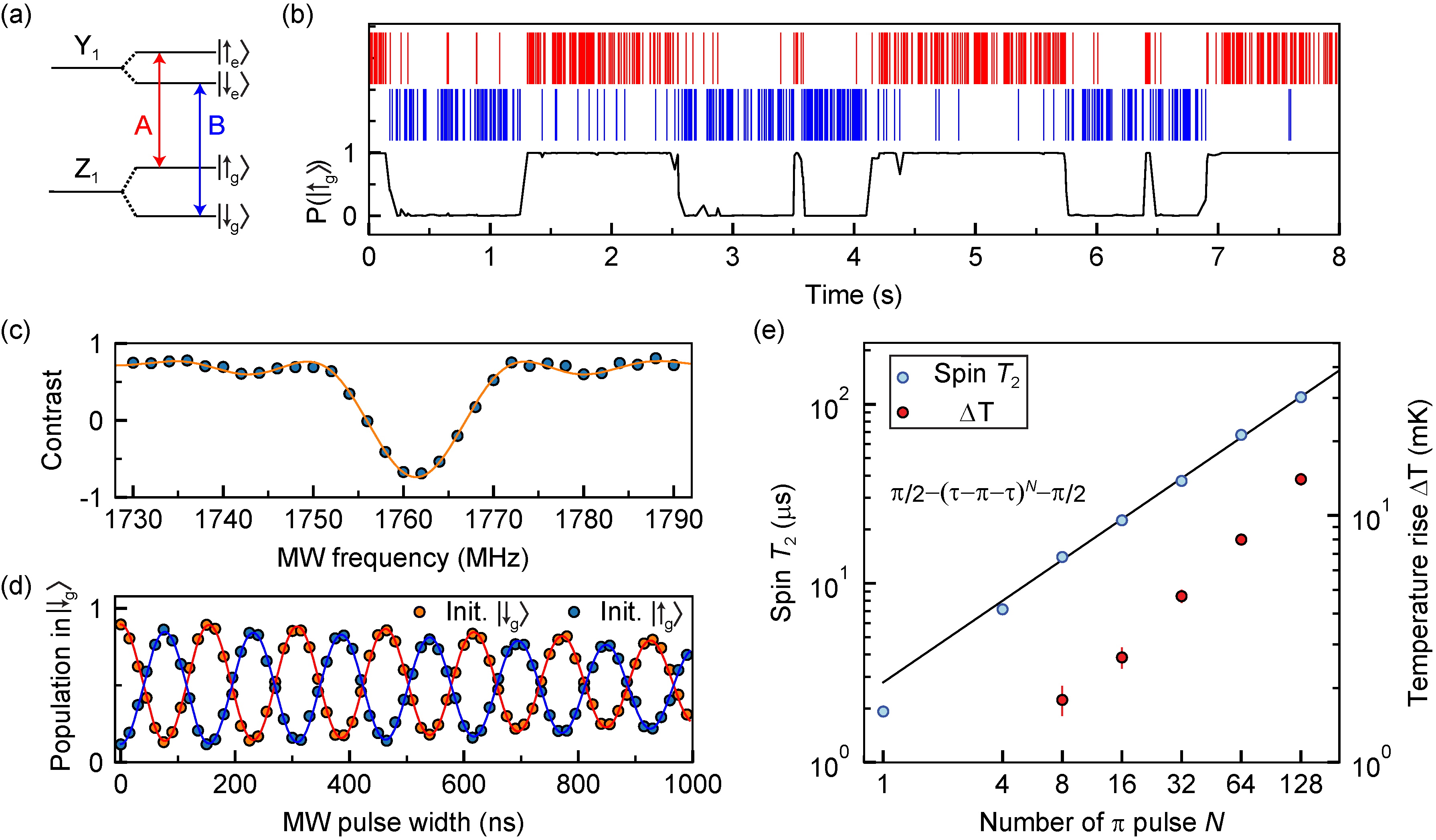}

    \caption{\textbf{Spin readout and coherent control of a single \er ion}.
    \textbf{(a)} Simplified level structure of \mbox{\er\!\!:YSO}.
    \textbf{(b)} Detected photons from (A, B) transitions are shown as red and blue lines respectively. Black curve shows the inferred probability of the ion to be in $\ket{\upg}$.
    \textbf{(c)} Ground state ODMR at ($B,\theta,\varphi$) = (112 G, 90$^\circ$, 130$^\circ$). The vertical axis is defined as the population contrast between $\ket{\downg}$ and $\ket{\upg}$. The orange line is a fit to the Bloch equation.
    \textbf{(d)} High-visibility Rabi oscillations can be driven between the ground spin states using a MW pulse with varying duration.
    \textbf{(e)} Scaling of spin $T_2$ (blue dots) with the number of $\pi$ pulses in the MW CPMG sequence. The black line is the power law fitting of the form $T_2 (N) = \text{A}\cdot N^{\alpha}$, with fitted A = $2.80 \pm 0.20$, $\alpha = 0.76 \pm 0.02$. Red dots show the cryostat heating while applying the MW sequences.
    } 
	\label{figure5}
	\end{figure*}

Achieving good coupling efficiency requires careful control of the fiber angles, which cannot be adjusted \emph{in situ}. We rely on precision machining to realize good angular alignment between the sample surface and the top surface of the Attocube mount, to which the PCB and quartz substrate are fixed. The fiber is glued down to the quartz CPW substrate with Stycast 2850FT (Catalyst 9), at the center of the two inner ground pads of the CPW (see Fig.~\ref{figure1}b). Top and side cameras are used to ensure correct alignment of the transverse positions and of the pitch and yaw angles during the gluing process. For the rotation angle alignment, we image a laser spot coming out of the angle-polished fiber via the top camera while changing the camera focus. We tune the rotation angle until the imaged laser spot only moves along the fiber axial direction when the camera focus is varied, indicating that the light from the fiber is propagating in a vertical plane.

By measuring the coupling efficiency between the same fiber and devices before gluing (in a six-axis mount) and after gluing (with only three-axis translation control), we observe an efficiency decrease about 2\%, indicating that the gluing procedure meets the required angle alignment tolerances. After cooling down the cryostat we observe another $\sim$4\% decrease in $\eta$ compared with the value measured at room temperature, suggesting potential fiber angle misalignment during the cooling process. The value of $\eta$ at cryogenic temperature does not change across repeated thermal cycles.

The complete unit can be aligned to a nanophotonic device with the help of a microscope with modest (10X) magnification. Furthermore, we have found that operation inside a cryostat with no optical access is possible using the following procedure. At room temperature, the fiber is aligned to the device using a temporary microscope suspended from the cryostat frame while the cryostat vacuum housing and radiation shields are removed. The coupling is verified by measuring the PC cavity resonance with a tunable laser (Fig.~\ref{figure2}g). Then, the fiber is backed away from the sample by $\sim$60 $\mu$m (measured using resistive encoders on the Attocube positioners) while maintaining some coupling. The microscope is then removed, and the cryostat is assembled, pumped out and cooled down. During the $\sim$14 hour cool-down, the translation stage is continuously adjusted (in the $\hat{x}$, $\hat{y}$-directions) to track the optimum coupling position. Once the base temperature is reached, the coupling is manually optimized in all three dimensions. The probe can be repeatably scanned over a relatively large field of devices using the encoder, which is more than accurate enough to locate the $\sim$10 $\mu$m-sized GCs across its entire scan range if their positions are known with respect to the initial device.

\section{Demonstration with single erbium ions}
\label{sec:usage} 

Lastly, we demonstrate the effectiveness of the scanning probe by performing single-shot spin state readout and coherent manipulation of a single \er ion in YSO \cite{raha2020}. In an appropriately-oriented magnetic field, there are two spin-conserving optical transitions originating from the two ground spin states (labeled A and B, Fig.~\ref{figure5}a). We perform a spin readout by alternately exciting these transitions. A time trace of recorded photons in a single run of experiment is shown in Fig.~\ref{figure5}b, which allows the spin state to be determined.

To demonstrate MW functionality, we study optically-detected magnetic resonance (ODMR) of the ground spin states (Fig.~\ref{figure5}c). High-visibility Rabi oscillations between ground spin states can be driven resonantly by a MW pulse with varying widths (Fig.~\ref{figure5}d). Furthermore, we demonstrate dynamical decoupling sequences with up to 128 $\pi$-pulses of duration 78~ns and peak power 17.8~W to extend the spin coherence time in the presence of nuclear spin noise. The longest sequence has a total MW pulse duration of approximately 10 $\mu$s, and repeats every 36~ms (limited by the readout duration) for an average duty cycle of $2.8\times10^{-4}$. The resulting cryostat heating, 17 mK, is consistent with the average MW power of 5~mW (Fig.~\ref{figure3}c).

By measuring the Rabi frequencies when driving ground spin states \cite{raha2020} at varying static magnetic field orientations (provided by separate magnetic coils), we can extract the parameters of the AC magnetic field generated by the CPW as ($B_\text{AC},~\theta,~\varphi$) = (3.0 G, 57.1$^\circ$, 0.8$^\circ$) for an input MW power of 1 Watt, in rough agreement with simulations. The small polar angle ($\theta$) discrepancy from the MW field simulations may come from the misalignment between CPW and PC cavity along the $\hat{x}$-direction (Fig.~\ref{figure1}a).

\section{CONCLUSION AND OUTLOOK}
\label{sec:conclusion} 

Several improvements to the design are possible. First, stronger AC magnetic field can be generated by moving the CPW closer to the sample surface, enabling faster MW driving with the same incident power. In recent experiments, we have found that replacing the terminus of the CPW with a raised wirebond can lead to more than three-fold increase in the Rabi frequency. Alternatively, it may be possible to recess the fiber using etched V-grooves in crystalline silicon substrates. Second, further improvement of the directionality of the GC (which is the dominant limitation on $\eta$) may be realized by attaching a thin high-index film (e.g. silicon) to the bottom of the fiber. Lastly, we note that this approach could also be deployed using alternative fiber-waveguide coupling techniques (\emph{ie}, adiabatic tapers or end-fire coupling) with higher demonstrated coupling efficiency.

In conclusion, we have demonstrated a MW-optical hybrid scanning probe capable of simultaneous optical and MW coupling to atomic defects in a nanophotonic cavity. With a single probe, we are able to address multiple devices inside a cryostat without any free-space optical access. The approach shown here may be readily generalized to other atomic defect spin systems.\vspace{0.5cm}

We thank Gerard Wysocki for a loan of fiber polishing equipment. Support for this research was provided by the National Science Foundation (NSF, EFRI ACQUIRE program Grant No.~1640959), the Princeton Center for Complex Materials (PCCM), an NSF MRSEC (DMR-1420541), the Air Force Office of Scientific Research (Grant No.~FA9550-18-1-0081), the DARPA DRINQS program (Grant No.~D18AC00015) and the Sloan Foundation. We acknowledge the use of Princeton's Imaging and Analysis Center, which is partially supported by PCCM, as well as the Princeton Micro-Nano Fabrication Lab and Quantum Device Nanofabrication Lab facilities. C.M.P.~was supported by the Department of Defense through the National Defense Science \& Engineering Graduate Fellowship Program.\vspace{0.5cm}

\textbf{\flushleft REFERENCES}
\bibliography{HybridProbe.bib}

\end{document}